\documentclass[12pt,a4paper]{article}

\usepackage[latin1]{inputenc}
\usepackage{amsmath}
\usepackage{amsfonts}
\usepackage{amssymb}
\usepackage{amsthm}
\usepackage{tabularx,ragged2e,booktabs,caption}
\usepackage{graphicx}

\usepackage[usenames]{color}

\usepackage{url}
\newcommand{\trace}{\hbox{\rm tr} }

\newcommand{\TT}{}

\input{my.sty}

\author{L.~ C.~ G.  Rogers\thanks{
Corresponding author: Statistical Laboratory, 
Wilberforce Road, Cambridge CB3 0WB, United Kingdom. 
\tt{lcgr1@cam.ac.uk}}
\\
P. Zaczkowski\thanks{
Statistical Laboratory, 
Wilberforce Road, Cambridge CB3 0WB, United Kingdom. 
}
\\
\\ Statistical Laboratory, University of Cambridge}
\title{Monte Carlo approximation to optimal investment}
\begin{document}
\maketitle

\bibliographystyle{acm}


\begin{abstract}
This paper sets up a methodology for 
approximately solving optimal investment
 problems using duality methods combined with Monte Carlo simulations.
  In particular, we show how to tackle high dimensional problems in 
  incomplete markets, where traditional methods fail due to the curse of dimensionality.
\end{abstract}


\section{Introduction.}\label{sec:introduction}
From the early work of Merton and his seminal papers \cite{Merton:1971} 
and \cite{Merton:1969}, the optimal investment literature has been
 trying to \TT{determine} how to invest  in financial
  markets when facing uncertainty. Over the  \TT{following 
  twenty} years, many general results \TT{were proved}, and many techniques for
    tackling the questions \TT{developed}. 

Deriving the abstract forms of the solutions is a great 
achievement of mathematical finance. However, anyone who wants
 to use them to guide them in making  investment decisions 
 will quickly find out that they are typically rather uninformative. 
 This is because, apart from a couple of highly stylized examples, 
 concrete numerical answers in optimal investment problems are simply
unobtainable, largely due to the curse of dimensionality. 
  See \cite{Rogers:2013} for a survey of the traditional methods 
  and a \TT{range} of examples where answers can actually be \TT{found}. 

The goal of this paper is to take a pragmatic approach. We take the 
point of view of an investor who is facing a particular market and 
is interested in knowing a \emph{good} thing to do at a particular
 time. Hence, we want to be able to describe what a good investment 
 strategy is in a particular market environment \emph{without}
  \TT{computing} the whole value function for the problem, \TT{and we
  want to quantify what we mean by a good investment strategy, in 
  terms of bounds on the objective.} 

Taking this standpoint lets us make progress by combining various 
optimization techniques that would fail individually when applied 
to difficult optimal investment problems. Namely, we shall use 
the Pontryagin--Lagrange approach to determine locally optimal 
trajectories; the dual formulation of the optimal investment 
problem to derive bounds on the optimal trajectory; and
 Monte Carlo techniques to approximate the expectation operator. 

Combining these related methods lets us handle a surprisingly large
 class of problems. We will show how to find approximately optimal
 investment paths for any continuous-path incomplete market driven 
 by a \TT{diffusion} factor process. As an illustration 
 of the effectiveness of the method, we shall provide a couple of 
 numerical examples. We shall start with the benchmark Merton 
 problem, moving on to problems that are increasingly more 
 difficult to handle numerically and mathematically. 

This paper is structured as follows. In Section \ref{sec:problem} we
present the general problem and the methodology for solving it. Section \ref{algos} describes the algorithms used in the method. In 
Section \ref{sec:numerics} we give numerical evidence for the 
performance of the method, considering examples of the Merton 
problem, the non-constant relative risk aversion, and finally 
a multi-dimensional incomplete market driven by a diffusion. 
Section \ref{sec:conclusions} concludes.
%
%
%


\section{Continuous markets driven by a diffusion.}\label{sec:problem}
We shall present the methodology in the context of a
finite-horizon optimal
investment-consumption problem where the volatilities and drifts of the
assets depend on some diffusion factor process. It will become evident that the
general approach is not limited to such examples, but it is easier to 
explain in this more concrete setting. We shall also make various assumptions
of boundedness on processes and global Lipschitz properties of coefficients
which could be relaxed, but which simplify the exposition and proof: the
aim is transparency, not maximality.

To begin with, suppose that $X$ is an $\R^k$-valued diffusion process
satisfying
\begin{equation}
dX_t = \sigma_X(X_t)\, dW_t + \mu_X(X_t) \, dt 
          \equiv \sigma_X\, dW_t + \mu_X \,dt,
\label{Xdyn}
\end{equation}
where $W$ is a $d$-dimensional Brownian motion, 
and $\sigma_X: \R^k \rightarrow 
\R^{k} \otimes \R^d$ and $\mu_X:  \R^k \rightarrow 
\R^{k}$ are globally Lipschitz coefficients. 

We shall consider an investor who is allowed to invest in a market with 
a riskless asset yielding interest at rate $r_t \equiv r(X_t)$, and $n$ 
stocks having volatility matrix $\sigma_t \equiv \sigma(X_t)$ and drift
 $\mu_t \equiv \mu(X_t)$.
  Here, $r: \R^k \rightarrow \R$, 
  $\sigma: \R^k \rightarrow
  \R^n \otimes \R^d$, and $\mu: \R^k \rightarrow 
  \R^n$ are bounded measurable functions. 
  We assume non-degeneracy of the market, that is,  $d \geq n$, 
  and that the row rank of $\sigma$ 
   equals $n$. When $n = d$, the matrix $\sigma$ is then invertible, 
   and we have a special case of a complete market.

With these assumptions in place, the investor's wealth $w_t$ at time $t$
evolves\footnote{We use the notations $a\cdot b$ for the 
scalar product of two vectors $a$ and $b$, and $\wun$ for the column
vector of ones.} as
\begin{align}\label{wdyn}
dw_t = r_t w_t\,   dt + \theta_t \cdot (\sigma_t \,dW_t + (\mu_t - r_t \wun)
\, dt) - c_t \,dt,
\end{align}
where the $n$-vector process
 $\theta_t$ represents the cash holdings in each of the stocks, 
 and $c_t$ denotes the agent's consumption rate.
The agent's objective at time $t$ is to achieve
\begin{align}\label{Vdef}
\sup_{(c,\theta) \in \sA} E \left[ \int_t^T U(s, c_s) ds + 
\varphi(w_T) \Big| w_t = w, X_t = x \right] \equiv V(t, w, x),
\end{align}
where $U$ and $\varphi$ are strictly concave $C^2$ utility functions satisfying 
the Inada conditions\footnote{These are the conditions
$ \lim_{c \downarrow 0} U_c(t, c) = \infty =\lim_{w \downarrow -\infty} \varphi'(w)$,
$ \lim_{c \uparrow \infty} U_c(t, c) =  0 =  \lim_{w \uparrow \infty} \varphi'(w)$.  }, and $\sA$ denotes the set of admissible 
consumption-portfolio pairs:
\begin{equation}
\sA = \{ (c,\theta):  \hbox{\rm $c$ and $\theta$ are previsible}, \;
c \geq 0, \hbox{\rm and for some $K<\infty$,} \; \| \theta_t \| \leq K  \}.
\label{sAdef}
\end{equation}

\medskip\noindent
{\sc Remarks.} (i)  Notice
that the function $\varphi$ is defined on the whole of $\R$.

\medskip\noindent
(ii) The above definition of admissibility \eqref{sAdef} 
is not the usual one\footnote{One typically imposes a non-negativity 
constraint on the wealth process associated with the trading 
strategy $\theta$.}. 
 We do not
expect that the supremum in \eqref{Vdef} will be attained within the 
set $\sA$, but as our goal is to come up with good sub-optimal strategies, 
this does not matter for our current purposes. Admissibility is imposed to eliminate doubling strategies, where wealth may go arbitrarily negative 
before time $T$, but ends up at a high value at time $T$. The assumptions
made here rule this out; if we  were to go  to large negative wealth at some
time in $(0,T)$, boundedness of $\sigma$, $\mu$ and $\theta$ prevent us 
returning to positive wealth with certainty by time $T$, and the penalty
imposed by the concave function $\varphi$ then makes this a bad thing to do.

\medskip\noindent
(iii) If the dimension $k$ of the statespace of the factor diffusion
$X$ were not very small, it is not feasible to calculate and store the
value function $V$. The approach we develop in this paper allows us to 
determine approximately optimal policies {\em without} the need to calculate
$V$.

\bigbreak
We shall require one technical condition on $U$, which is expressed as 
a condition on the inverse marginal utility $I$, defined by
\begin{equation}
U_c(s, I(s,z) ) = z,  \qquad\qquad (z>0).
\label{Idef}
\end{equation}
We require
\begin{equation}
\hbox{ {\sc Assumption:} there exists $\alpha, \; A>0$ 
such that $I(t,z) \leq A(1+z^{-\alpha})$ }. 
\label{assumption}
\end{equation}
The inequality has to hold for all $z>0$ and all $t \in [0,T].$

\bigbreak
We are now ready to state the main result of the paper, which allows us
to derive effective Monte Carlo bounds on the value, and to find 
good sub-optimal strategies  {\em pathwise}.  The proof uses 
duality arguments similar to those presented in  \cite{Cox:1989},
 \cite{Karatzas:1987}, and later described in a more general setting in 
 \cite{Karatzas:1989}. 

\begin{theorem}\label{thm1}
Suppose that $\kappa$ is a bounded previsible process such that 
\begin{equation}
\mu_t - r_t \wun - \sigma_t \kappa_t = 0,
\label{kappa}
\end{equation}
and that $\zeta$ solves the linear SDE
\begin{equation}
d\zeta_t = \zeta_t ( -\kappa_t \, dW_t - r_t \, dt).
\label{zeta}
\end{equation}
Define the function $g$ by\footnote{
The functions $\tilde U$, $\tilde \varphi$ are the convex
dual functions, 
$\tilde U(t,z) \equiv \sup_x\{ U(t,x) -zx\}$, $\tilde\varphi(z)
\equiv \sup_x \{ \varphi(x) - zx\}$.
}
\begin{equation}
g(t,z,x) = E \Big[ \int_t^T \tilde{U}(s, \zeta_s) ds
 + \tilde{\varphi}(\zeta_T)\;  \Big | \; \zeta_t = z, X_t = x \; \Big]
 \label{gdef}
\end{equation}
for $ t \in [0,T]$, $z>0$, $x \in \R^k$.
Then for any $t \in [0,T]$, $z>0$, $w\in\R$, $x \in \R^k$, 
and bounded previsible $\theta$, we have the inequalities
\setlength{\fboxsep}{15pt}
\begin{equation}\label{bounds}
\framebox{$g(t, z, x) + w z - h(t, w,z, x, \theta )
 \leq V(t, w, x) \leq g(t, z, x) + w z,$}
\end{equation}
where
\begin{equation}\label{hdef}
h(t, w,z, x, \theta ) \equiv \mathbb{E} \Big[ 
\; \tilde{\varphi}(\zeta_T)-
\varphi(w_T^{{\theta}}) + \zeta_T\,  w_T^{{\theta}}  
\; \Big| \; w_t = w,  \zeta_t = z, X_t = x \;\Big] ,
\end{equation}
and the process $w^\theta$ is the solution to the wealth
evolution \eqref{wdyn} with portfolio process $\theta$ and 
consumption process 
\begin{equation}
c_s = I(s,\zeta_s), \qquad\qquad (s \geq t).
\label{copt}
\end{equation}
\end{theorem}

\bigbreak
\noindent
{\sc Remarks.} (i) In general the matrix $\sigma$ is not even 
square, so not invertible, but we could try to find $\kappa$ to 
satisfy \eqref{kappa} by taking the pseudo-inverse of $\sigma$:
\begin{equation}
\kappa_t = \sigma_t^T (\sigma_t\sigma_t^T)^{-1}(\mu_t - r_t \wun).
\label{kappa_pi}
\end{equation}
This can be done if $(\sigma_t\sigma_t^T)^{-1}$ is bounded, in effect
a uniform ellipticity condition of the kind commonly imposed in
such problems.

\medskip\noindent
(ii) From the definition of the convex dual function $\tilde\varphi$, 
it is clear that $h$ is always non-negative. Since $h$ dominates the
gap between the lower and upper bounds, we should aim to make $h$ as
small as we can.  Ideally, we would have that $h$ was zero, which would
require us to have
\begin{equation}
\varphi'(w_T) = \zeta_T.
\label{transversality}
\end{equation}
If we demanded that this happens, then the problem becomes a BSDE with 
\eqref{transversality} as the terminal condition. As it seems that there
are as yet no efficient numerical methods for solving BSDEs in high
dimensions, this does not help much.  What we are attempting to do with
this approach is in effect relax the demand that the solution we construct
hits the terminal condition \eqref{transversality}, but instead to estimate
the error we make when we fail to match the terminal condition.

\bigbreak\noindent
{\sc Proof. (a) The upper bound.}
The process $\zeta$ is determined by \eqref{kappa} and \eqref{zeta};
in what follows, we shall suppose that $c$ is determined from $\zeta$
by \eqref{copt}.

Consider the It\^{o} expansion of $\zeta_T w_T$. We have:
\begin{eqnarray}
0 &=& - \zeta_T w_T + \zeta_t w_t + \int_t^T 
          (\zeta_s dw_s + w_s d\zeta_s + d[\zeta, w]_s)
          \nonumber
\\
&=& - \zeta_T w_T + \zeta_t w_t + \int_t^T \zeta_s(
\theta_s \cdot \sigma_s - w_s\kappa_s) \; dW_s
\nonumber
\\
\nonumber
&&\qquad +\int_t^T \zeta_s( r_s w_s  + \theta_s \cdot (\mu_s -r_s \wun) - c_s
-r_s w_s  -\theta_s \cdot\sigma_s \kappa_s
)\; ds
\\
&=&- \zeta_T w_T + \zeta_t w_t + \int_t^T \zeta_s(
\theta_s \cdot \sigma_s - w_s\kappa_s) \; dW_s - \int_t^T \zeta_s c_s \; ds
\end{eqnarray}
using \eqref{kappa} and \eqref{zeta}.

We claim that the stochastic integral has zero mean, and in order
to establish this, it is necessary to control the integrand. The
processes $\kappa$, $\theta$, and $\sigma$ are all bounded
by hypothesis, so we need to have control on $\zeta$ and $w$.
Since $\zeta$ satisfies the linear SDE \eqref{zeta}
with bounded coefficients $\kappa$ and $r$, it is not hard to 
establish a bound on $E[ (\zeta^*_t)^p]$ for any $t>0$, 
and for any $ p \geq 2$, where
$\zeta^*_t \equiv \sup_{0 \leq s \leq t} |\zeta_s|$; see, for
example, Lemma V.11.5 of \cite{RW2}. Similarly, 
we may bound $E[ (\zeta^*_t)^{-p}]$ for any $t>0$, 
and for any $ p \geq 2$, by considering the linear SDE for
$\zeta^{-1}$.
 All that remains is to 
establish a similar bound for $w^*_t$, where $w$ is given by
\eqref{wdyn}. The only problematic part of this estimation is 
in controlling $c$, but this is where the Assumption \eqref{assumption}
comes in, since $\zeta^{-1}_t$ is controlled as before, and $c$ is 
bounded by some power of $\zeta$.  

We therefore conclude that
\begin{equation}
0 = E \biggl[\;
- \zeta_T w_T + \zeta_t w_t
 - \int_t^T \zeta_s c_s \; ds
\;\biggr].
\label{zero}
\end{equation}
We can add this equality to \eqref{Vdef} to find\footnote{We use 
\eqref{copt} at the first step.}
\begin{eqnarray}
V(t,w,x) &=& \sup_{(c,\theta) \in \sA} E \biggl[ \int_t^T
\{ U(s, c_s) - \zeta_s c_s \}\; ds + 
\varphi(w_T)  - \zeta_T w_T  +
\nonumber
\\  &&
 \qquad\qquad\qquad\qquad\qquad
 + \zeta_t w_t
\Big| w_t = w, X_t = x,\zeta_t = \zeta \biggr]
\label{ub0}
\\
&\leq & E \left[ \int_t^T
\tilde U(s, \zeta_s)\; ds + 
\tilde\varphi(\zeta_T)   + \zeta_t w_t
\Big| w_t = w, X_t = x,\zeta_t = \zeta \right]
\nonumber
\\
&=& E \left[ \int_t^T
\tilde U(s, \zeta_s)\; ds + 
\tilde\varphi(\zeta_T)  
\Big| w_t = w, X_t = x,\zeta_t = \zeta \right] + \zeta w
\nonumber
\\
&=& g(t,\zeta,x) + \zeta w.
\label{ub1}
\end{eqnarray}
This is the upper bound in \eqref{bounds}.

\medskip\noindent
{\sc (B) The lower bound.} The argument reuses elements of the proof
of the upper bound. The task this time is to propose some admissible
$(c,\theta)$ and deduce a lower bound from it.

 Given the state-price density process $\zeta$
as in \eqref{zeta}, our intention is to use the process $c$ to be defined
from it by \eqref{copt}. Doing this, we see that the integral term
appearing in the right-hand side of  
\eqref{ub0} is equal to 
\begin{equation}
E \int_t^T \tilde U(s, \zeta_s) \; ds,
\nonumber
\end{equation}
and moreover that \eqref{zero} still holds by the same argument as
before.
For any bounded previsible $\theta$, the pair $(c,\theta)$ is admissible, 
so if we use that admissible pair we find as at \eqref{ub0} that
\begin{eqnarray}
V(t,w,x) &\geq& E \biggl[ \int_t^T
\tilde U(s, \zeta_s)\; ds +
\varphi(w^\theta_T)  - \zeta_T w^\theta_T   + \zeta_t w^\theta_t 
\Big| w_t = w, X_t = x,\zeta_t = \zeta \bigr]
\nonumber
\\
&=& g(t,\zeta,x) + w \zeta - h(t,w,\zeta,x,\theta)
\label{lb0}
\end{eqnarray}
when we recall the definitions \eqref{gdef} and \eqref{hdef} of 
$g$ and $h$.

\hfill$\square$

\bigbreak\noindent
{\sc Remarks.} (i) For any bounded previsible $\theta$ and $\kappa$
the result \eqref{bounds} of Theorem \ref{thm1} gives two-sided bounds
on the value function. Importantly, the numerical values of $g$ and
$h$ can be estimated {\em by forward simulation from current values}.
It is also worth noting that the methodology does not require any
`simulation within simulations' which substantially increases the
computation times; we will be evaluating the state-price density and
the portfolio process along {\em just one trajectory.}
All we need to do is to simulate 
  sufficiently many sample paths to approximate the 
  expectation operator in \eqref{gdef} and \eqref{hdef}.
\\
(ii)  We need to have a measure for comparison
  between the bounds in \eqref{bounds}. Since utility 
  functionals are defined up to affine transformations, our 
  measure needs to be invariant under those. Thus 
  the difference between the upper and lower bounds is not 
  informative. 
  
  We can however think of giving up a fraction of the initial wealth $\alpha w$ 
and look for the minimal $\alpha$ such that the upper bound corresponding to
 $(1-\alpha)w$ initial wealth is at most as large as the lower bound for 
 starting with wealth $w$. This $\alpha$ is of course:
 \begin{equation}
 \alpha(t, w, \zeta, X, {\theta}) \equiv 
 \frac{h(t, w, \zeta, X, {\theta})}{\zeta w},
 \label{error}
 \end{equation}
which will from now on be our efficiency measure. Notice that \eqref{error} is
 a dimensionless quantity.
 \\
(iii)  The key issue for obtaining good bounds is of course the 
choice of the processes $\kappa$ and $\theta$.
The traditional way to approach solving the problem \eqref{Vdef} would 
be to write down the HJB equation, derive the corresponding PDEs,
 and try to solve them. However, these PDEs are typically highly 
 non-linear, and we only stand a chance of getting reasonably 
 stable solutions in dimensions one or two. 

Nevertheless, we can deduce some worthwhile information from the HJB equation.
 Dropping the $t$ subscript, and remembering the function $V$ 
 takes $(t, w, X)$ as arguments, the HJB equation is
 \begin{eqnarray}
 0 &=&  \sup_{c, \theta} \bigl[\; U(t, c) + V_t + \left( rw + \theta
 \cdot(\mu - r\wun)
 - c\right)V_w + \mu_X \cdot V_X + 
 \nonumber
 \\
 &&\qquad\qquad  
 + \frac{1}{2} |\theta^T \sigma|^2 V_{ww}
 + \theta\cdot\sigma\sigma_X^T\, V_{Xw} + \half\trace(\sigma_X\sigma_X^T V_{XX})\;\bigr].
 \label{hjb1}
 \end{eqnarray}
Optimizing over $c$ leads to the conclusion that $c_t = I(t,V_w)$, and 
optimizing over $\theta$ tells us that we should have
\begin{equation}
\theta = - (\sigma \sigma^T)^{-1} \bigl\lbrace\;
(\mu - r\wun) V_w + \sigma\sigma_X^T \,V_{Xw}
\;\bigr\rbrace/V_{ww}.
\label{theta_star}
\end{equation}
Here $\sigma \sigma^T$ is invertible by our non-degeneracy assumptions on the market.

Assuming that $V$ and $g$ are dual (as we would expect from \eqref{bounds}), in that
\begin{equation}
V(t,w,x) = \inf_\zeta\{ g(t,\zeta,x)+w\zeta\},
\qquad g(t,\zeta,x) = \sup_w\{ V(t,w,x)-w\zeta\},
\label{dual0}
\end{equation}
this would lead us to the relations
\begin{equation}
w = -g_z(t,z,x), \qquad \zeta = V_w(t,w,x).
\label{dual1}
\end{equation}
Straightforward calculus then leads to
\begin{align}\label{Vww}
V_{ww}(t, w, x) = -1/g_{\zeta \zeta}(t, \zeta, x).
\end{align}
These relations help us to make choices of $\kappa$ and $\theta$. 
We will use \eqref{kappa_pi} to make our (pathwise) choice for $\kappa$, 
and then we will use the truncated form
\begin{equation}
\theta = - (\sigma \sigma^T)^{-1}(\mu - r\wun) V_w  /V_{ww}
= (\sigma \sigma^T)^{-1} (\mu - r\wun) \,\zeta g_{\zeta \zeta}(t, \zeta, X)
\label{theta1}
\end{equation}
for the pathwise choice of $\theta$. We should in principle include the 
cross derivative term 
from \eqref{theta_star}  in the choice of $\theta$, and in some situations it
might well be worth doing this, but the cost is that we have to get hold
of the derivative of $\zeta$ with respect to $X$, and doing this by simulation
is cumbersome. The virtue of the form \eqref{theta1} is that we just need
the second derivative of the convex function $g$ with respect to its scalar
argument $\zeta$, and determining this by simulation is computationally 
feasible.

\medskip\noindent
(iv) In practice, it will be clumsy to form an estimate of the term
$h$ in \eqref{bounds} if we are determining the portfolio process
$\theta$ according to the recipe just outlined, because if we are to 
simulate an evolution of $(X,w)$ we will at each step need to 
identify derivatives of $g$, and this is a simulation within a 
simulation.
We envisage the 
lower bound in \eqref{bounds} being used as a means to {\em assess}
a particular portfolio rule which may be expressed explicitly as
some function of $(t,X,w)$. In a high-dimensional problem,
we do not expect the optimal portfolio rule to be something we can
characterize, but we may well have some heuristic for some `good'
portfolio rule, and \eqref{bounds} gives us a way to tell how
good that heuristic may be.

\vspace{3mm}

\noindent \textbf{Summarising:} Given an initial state $(t, w, \zeta)$, we can 
follow the dynamics of $w$, $\zeta$, and $X$, using 
\eqref{Xdyn}, \eqref{wdyn}, \eqref{zeta},
$\kappa$ given by \eqref{kappa},
$c$ given by \eqref{copt}, and $\theta$ given by \eqref{theta1} (or perhaps
\eqref{theta_star}).

The key advantage of this formulation is that all we need to do now is to optimise 
the bounds \eqref{bounds} for a one-dimensional starting value of the dual 
process $\zeta$. This is a quick procedure numerically.

\section{Algorithms.}\label{algos}

We will now describe an algorithm for simulating the optimal path and controls for 
the problem \eqref{Vdef}, given \emph{a particular realisation of the Brownian motion}.
 That is, we do not attempt to recover the whole value function, as this is bound to 
 fail in higher dimensions. Our method, which is effectively local, will follow a
  particular realisation of the Brownian motion $W$ and tell us how to invest 
  and consume in that particular case. After all, one is predominantly interested 
  in how to invest in the current market conditions, and does not necessarily care
   about all possible versions of reality!

\vspace{3mm}
\renewcommand{\tablename}{Algorithm}
\begin{table}
\caption{Computing the optimal path.} 
\begin{tabular}{ l p{14cm}}
\hline
\hline               
\emph{Step 1:}  & \textbf{Initialisation.} Pick starting values $w = w_0$, $X = X_0$
 and a grid of time steps $0 = t_0 < t_1 < t_2 < \dots < t_N = T$ along which we want 
 to know the solution. Simulate a realisation of the Brownian motion $W$ along which 
 we want to calculate the optimal path. \\
\emph{Step 2:} & \textbf{Finding the optimal $\zeta_0$.} For any $\zeta$, we can 
calculate $g(0, \zeta, X) + w \zeta_0$. This function is convex in $\zeta$, so we
 can use the golden Section search to find the minimum in \eqref{dual0}. This gives 
 us the value of $V(0, w_0, X_0)$ and the optimal starting value of the dual process $\zeta_0$. \\
\emph{Step 3:} & \textbf{Calculating the optimal path.} For each $n = 0, 1, \dots, N-1$,
 we have $(t_n, \zeta_{t_n}, X_{t_n})$ available. We use \eqref{copt} to work 
 out $c_{t_n}$, \eqref{theta1} to work out $\theta_{t_n}$, and \eqref{dual1} 
 to work out $w_{t_n}$. We then cacluate $\kappa_{t_n}$  wth \eqref{kappa} and use
  the Euler scheme to move to time $t_{n+1}$ using \eqref{Xdyn} and \eqref{zeta}. \\
  \hline
  \hline  
\end{tabular}\label{algo1}
\end{table} 
\renewcommand{\tablename}{Table} 

\vspace{3mm}


Algorithm \ref{algo1} describes how to compute the best bounds numerically. 
The cost of running this algorithm will be $\mathcal{O}(N) \times \mathcal{O}(g)$,
 where $\mathcal{O}(g)$ is the average cost of evaluation of the function $g$ and $h$. 

In Algorithm \ref{algo1}, we have not yet given the details of how to calculate
 the function $g$ numerically (which
 will be the business of Algorithm \ref{algo2}).
  That is, we want to be able to numerically calculate 
 the expectation in \eqref{gdef} and \eqref{hdef} for $t = t_n$, being one of 
 the grid points in the time discretization. We approach the calculation 
 numerically with Monte Carlo methods, sampling $M$ paths of Brownian 
 motion $W$ for $t = t_n, t_{n+1}, \dots, t_N$, simulating the values 
 of the functional in the expectation of \eqref{gdef} and \eqref{hdef}, 
 and finally averaging over the sampled paths.

In practice, we find that it might be necessary to use importance sampling
 in order to decrease the volatility of our estimates. In order to do that, 
 define the change of measure martingale
\begin{align}\label{Zdyn}
dZ^{-1}_s = Z_s^{-1} \sigma^Z_s dW_s \text{ for } t \leq s \leq T, \qquad Z_t = 1,
\end{align}
and set $\frac{d \mathbb{Q}}{d \mathbb{P}} |_{\mathcal{F}_t} = Z_t^{-1}$. 
Then we can rewrite \eqref{gdef} and \eqref{hdef} as
\begin{align}\label{greduced}
g(t, \zeta, X) & = \mathbb{E}^\mathbb{Q} \left[\int_t^T Z_s \tilde{U}(s, \zeta_s) ds
 + Z_T \tilde{\varphi}(\zeta_T) \Big| \zeta_t = \zeta, X_t = X \right],
 \\
h(t, w, \zeta, X,{\theta}) & = \mathbb{E}^\mathbb{Q} \Big[
Z_T \varphi( w_T^{{\theta}}) - Z_T w_T^{{\theta}} \zeta_T - 
 Z_T \tilde{\varphi}(\zeta_T)\Big| w_t = w,  
\zeta_t = \zeta, X_t = X \Big] .
\end{align}
with a new Brownian motion $W^{\mathbb{Q}}$ under $\mathbb{Q}$ defined by
\begin{align}\label{WQdef}
d \bar{W}_t = dW_t - \sigma^Z_t dt.
\end{align}
The idea now is to choose $\sigma_Z$ in a way that the Ito expansion of the 
term $Z_T \tilde{\varphi}(\zeta_T)$ has no $d\bar{W}$ term. 
This has a variance reducing property. Writing $\dot{=}$ whenever two 
sides of an equality differ only by integrals with respect to $ds$, we have
\begin{align}\label{variancereduction}
Z_T \tilde{\varphi}(\zeta_T) & = Z_t \tilde{\varphi}(\zeta_t) + \int_t^T
 d(Z_s d\tilde{\varphi}(\zeta_s)) \dot{=}  Z_t \tilde{\varphi}(\zeta_t)  + \int_t^T \left( Z_s  
\tilde{\varphi}'(\zeta_s) d\zeta_s +  dZ_s \tilde{\varphi}(\zeta_s) \right) \\
 & = Z_t \tilde{\varphi}(\zeta_t) + \int_t^T Z_s \left( -\kappa_s
   \tilde{\varphi}'(\zeta_s) \zeta_s - \sigma^Z_s  \tilde{\varphi}(\zeta_s) \right)d\bar{W}_s.
\end{align}
Therefore, we set:
\begin{align}\label{sigmaZdef}
\sigma_Z \equiv - \kappa_s \frac{\zeta_s  \tilde{\varphi}'(\zeta_s) }
{ \tilde{\varphi}(\zeta_s) },
\end{align}
which cancels the $d\bar{W}$ term in \eqref{variancereduction}, and in turn in \eqref{greduced}.

With this in mind, we now present the numerical algorithm for calculating $g(t, \zeta, X)$. 

\vspace{3mm}


\renewcommand{\tablename}{Algorithm}
\begin{table}
\caption{Computing $g(t_n, \zeta, X)$ and $h(t_n, w, \zeta, X, \tilde{\theta})$} 
\begin{tabular}{ l p{14cm}}
\hline
\hline               
\emph{Step 1:}  & \textbf{Initialisation.} Recall $t = t_n$. Generate $M$ paths 
of Brownian motion $\bar{W}^i_t$, $i = 1, 2, \dots, M$, with values evaluated
 at $t = t_n, t_{n+1}, \dots, t_N$. The corresponding paths for $\zeta$, $X$,
  $w$ and $Z$ are denoted by $\zeta^i$, $X^i$, $w^i$ and $Z^i$ with 
  $\zeta^i_{t_n} = \zeta$, $X^i_{t_n} = X$, $Z^i_{t_n} = 1$ and $w^i_{t_n} = w$. \\
\emph{Step 2:} & \textbf{Simulation.} For $k = n, n+1, \dots, N-1$, update
 $\zeta^i_{t_{k+1}}$, $X^i_{t_{k+1}} $, $Z^i_{t_{k+1}}$ and $w^i_{t_{k+1}}$ 
 as follows. Equations \eqref{sigmaZdef} and \eqref{WQdef} give us the
  corresponding $dW_{t_k}$. We then use \eqref{zeta}, \eqref{Xdyn}, 
  \eqref{Zdyn} and \eqref{wdyn} to move to the next time point using the Euler scheme. 
   \\
\emph{Step 3:} & \textbf{Averaging.} Having calculated paths $\zeta^i$, $X^i$, $Z^i$ 
and $w^i$ corresponding to $M$ paths of $\bar{W}^i$, we return the approximate
 values of $g$ and $h$:
\begin{align}
g(t_n, \zeta, X)   \approx \frac{1}{M} \sum_{i=1}^M \left( \sum_{k=n}^{N-1} 
Z^i_{t_k} \tilde{U}(t_k, \zeta^i_{t_k}) + Z_{t_N}^i \tilde{\varphi}(\zeta_{t_N}^i) \right)
 \label{gdiscretized}
  \\
h(t_n, w, \zeta, X, \tilde{\theta})  \approx \frac{1}{M} \sum_{i = 1}^M Z^i_{t_N}
 \left(\varphi(w_{t_N}^i) - w_{t_N}^i \zeta_{t_N}^i - \tilde{\varphi}(\zeta_{t_N}^i) \right).
\end{align} \\
  \hline
  \hline  
\end{tabular} \label{algo2}
\end{table}
\renewcommand{\tablename}{Table} 

\vspace{3mm}


The computational complexity of Algorithm \ref{algo2} comes from 
\eqref{gdiscretized}, where we clearly see that we need 
$\mathcal{O}(N) \times \mathcal{O}(M)$ operations. Therefore, 
we deduce that $\mathcal{O}(g) = \mathcal{O}(MN)$.

The key to performance of the method is of course the accuracy of the Monte Carlo 
simulation. As we shall see in the following Section, the numerical results are 
promising. Even a fairly moderate number of Monte Carlo paths can provide a good 
approximation to the true value of $g$ and $h$. With this in mind, we proceed to
 examine the numerical results for the performance of the method.

\vspace{3mm}



\section{Numerical performance}\label{sec:numerics}
In this Section, we shall compare the results of the Monte Carlo solutions with
 special cases of the problem \eqref{Vdef} where we either know the solution 
 in closed form, or we know highly accurate numerical schemes for approximating the solution. 


We start off by analysing complete markets where some of the analysis in the 
previous Section simplifies. Recall that, in a complete market 
the asset volatility matrix $\sigma$ is invertible. 
This means we have a unique\footnote{Up to a multiplicative constant still to
 be found.} state-price density for the problem, given by
\begin{align}
\zeta_t = \zeta_0 \exp\left[ - \int_0^t \kappa_s\cdot dW_s - \int_0^t \left( r_s + \frac{1}{2} | \kappa_s |^2 \right)ds \right],
\end{align}
where $\kappa_s \equiv \sigma_s^{-1} (\mu_s - r_s 1)$.  Therefore,
  provided that \eqref{dual0} holds, our Monte Carlo method should be able to find 
  the optimal path exactly, modulo numerical errors coming from Monte Carlo
   approximation of the expectation operator in \eqref{gdef}, approximating 
   the derivatives in \eqref{dual1} and \eqref{Vww}, and finally the numerical 
   optimisation over the (scalar!) value $\zeta$ in \eqref{gdef}. The positive 
   side is that all these errors can be made small provided we use enough computational power. 

With that in mind, we start off with two examples of problems dealing with complete markets
 where the benchmark answers are reliable; and finish by analysing runs in incomplete 
 markets where we provide estimate error bounds, but where 
 no other solutions methods are available. 

\subsection{The Merton problem}\label{Merton}
We start by  comparing our results to the solutions of the Merton problem, which 
are available in closed form in multiple dimensions. Recall that the Merton problem 
assumes that functions $r$, $\mu$ and $\sigma$ in \eqref{wdyn} are constant, and the
 utility functions $U$ and $\varphi$ in \eqref{Vdef} take a particular form:
\begin{align}
U(t, c) & =  e^{-\rho t} u(c), \\
\varphi(w) & = A u(w),
\end{align}
where $a, b, \rho$ are positive constant, and $u$ is a
constant relative risk aversion utility:
\begin{align}
u(c) = \frac{c^{1- R}}{1-R},
\end{align}
for $R>0$, $R \neq 1$. Then the optimal solution takes the form:
\begin{align}
V(t, w, X) & = f(t) u(w), \\ 
\theta_t & = \pi_M w_t, \\
c_t &= \gamma(t) w_t,
\end{align}
where
\begin{align}
f(t) & = \left\{ A^{1/ R} e^{-b(T-t) } + 
\frac{e^{-\rho t/R}}{b+\rho /R} 
 ( 1 - e^{-(b+\rho/R)(T-t) }) \right\}^{R} ,\\
\pi_M & = R^{-1} (\sigma \sigma^T)^{-1} (\mu - r \wun), \\
\gamma(t) & =  e^{-\rho t / R} f(t) ^{-1/R},
\end{align}
where $b = (R-1)(r + |\kappa|^2/2R)/R$; see \cite{Rogers:2013}, Section 2.1.

\begin{figure}
\begin{center}
\includegraphics [angle=90,height= 20cm, width=15cm]{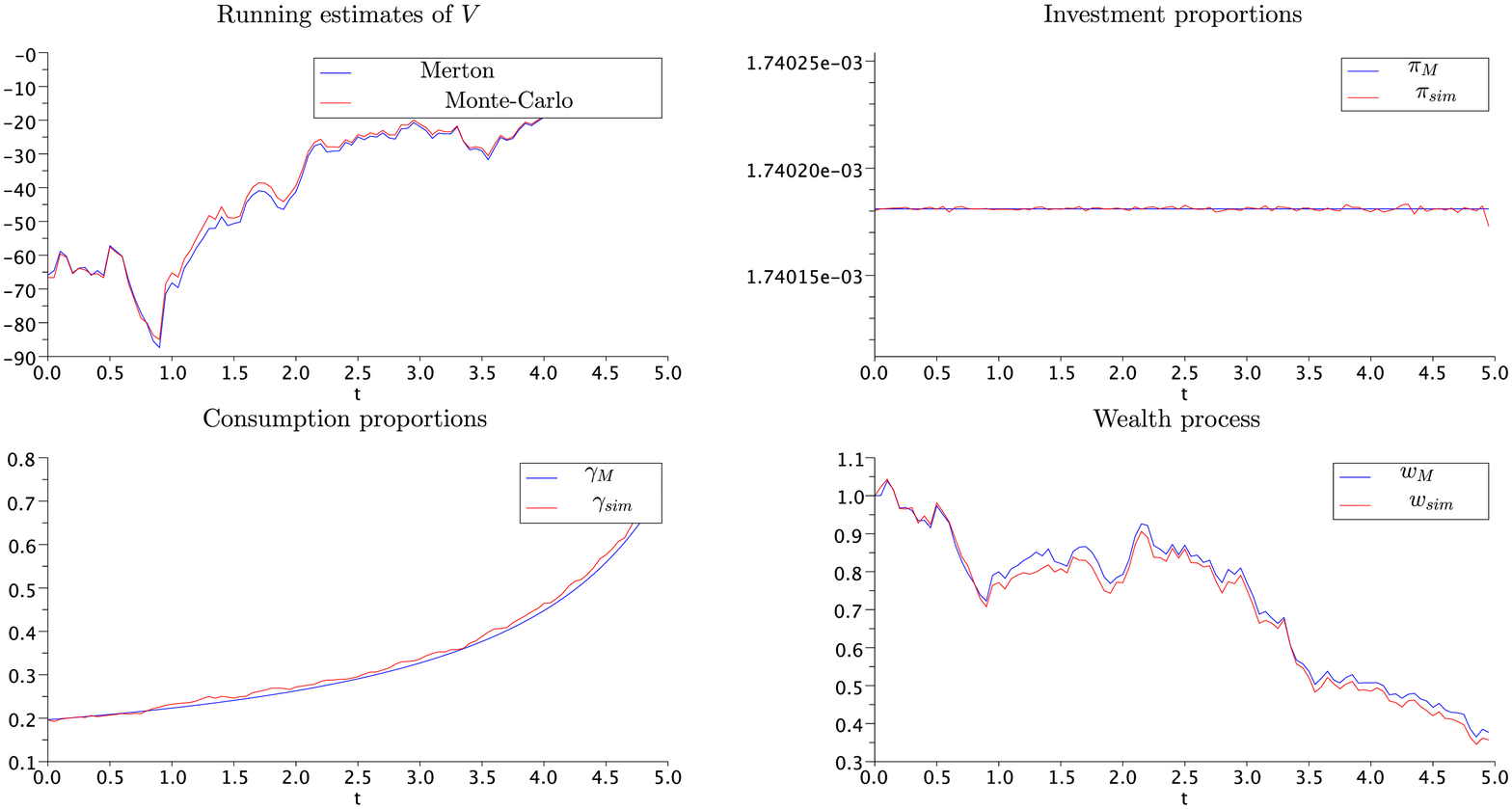}
\caption{\textbf{ Monte Carlo solution to the Merton problem.}
Here we
take $k = 3$, $r = 0.05$, $\rho = 0.03$, $R=3$, $a = 1$, $b = 2$, $N = 100$,
 $dt = 0.05$, $M = 1000$, $w_0 = 1$, $\mu = [0.07; 0.25; 0.15]$, $\sigma = 
 [0.12, 0.01, 0.03; 0.01, 0.50, 0.01; 0.03, 0.01, 0.27]$.}
\label{fig:Example1}
\end{center}
\end{figure}

Figure \ref{fig:Example1} shows the results of the simulation runs 
for the 3-dimensional version of the problem using $M = 1000$ paths.
 The top left panel shows the running estimate 
of the value function $V_M(t, w_M(t))$ along a particular realization
 of Brownian motion $W$. 
The top right and bottom left panels show investment and consumption proportions,
 respectively. Finally, the bottom right panel depicts the estimated wealth process 
 compared to the Merton wealth process. 

As we see, all the graphs give a very satisfactory approximation to the Merton solution. 
This is especially remarkable taking into account that we are already in dimension $3$, 
and we have used relatively few paths. 

We now present the study of how the accuracy of the solutions to the Merton problem
 varies for different values of the number of simulations $M$ and number of dimensions
  $K$. We found that the number of time steps $N$ used to discretize the integral 
  in \eqref{greduced} does not greatly influence the accuracy of the solutions. 

We compare the estimates of the optimal starting $\zeta_0$ found by the procedure 
\eqref{dual0} in Algorithm \ref{algo1}. For each test, we keep the initial data 
of \emph{Step $1$} fixed. We then run \emph{Step $2$} of Algorithm \ref{algo1}, each 
time approximating the function $g$ with a different set of Monte Carlo paths. 
This way, we can investigate how sensitive our optimized values of $\zeta_0$ are 
to the Monte Carlo procedure for approximating the expectation operator.

Table \ref{table3} and Table \ref{table4} present the results of the simulations for
 different number of Monte Carlo paths to calculate $g$, $M = 1000$ and $M = 10000$, 
 respectively. We see that the numerical results work reasonably 
 well for $K \leq 6$ when we choose to use $1000$ Monte Carlo paths. The average 
 $\zeta_0$ is pretty close to the true value, and the volatility of the estimates 
 stays modest. However, for larger values of $K$, we see that the estimates are 
 either not as accurate, or become more volatile. 

For $M=10000$, the results look much better. For $K \leq 9$, we see a considerable 
drop in the volatility of the estimates, and all of them lie within two standard 
deviations of the true value, with most of them being less than one 
standard deviation away. 

These results are very encouraging. They show that, even in dimensions up to $10$, having
 a reasonably modest number of Monte Carlo paths of $10000$ can provide satisfactory 
 results when solving the Merton problem. This is particularly interesting since the
  traditional HJB approach would struggle in these dimensions unless the problem has 
  a particular structure such that we can work out the value function explicitly. 

One might think that the accuracy of the method relies on the special structure of
 the Merton problem. We now show that this is not the case. We consider departures 
 from the basic problem where accurate numerical solutions are available.


\begin{table}
\begin{center}
{\footnotesize
\begin{tabular}{ | c | c | c | c | c | c | c | c | c | c | c | }
\hline
& $K = 1$ & $K =2$ & $K =3 $ & $K =4$ & $K =5$ & $K =6$ & $K =7$ & $K =8$ & $K =9$ & $K =10$ \\
\hline
Merton & 9.97 & 9.49 & 8.92 & 8.61 & 8.17 & 8.02 & 7.73 & 7.44 & 7.11 & 6.68\\
\hline
Average($\zeta_0$) & 9.72 & 9.33 & 8.64 & 8.86 & 7.53 & 7.85 & 7.36 & 7.54 & 6.44 & 5.31 \\
\hline
Stdev($\zeta_0$) & 0.12 & 0.14 & 0.23 & 0.34 & 0.30 & 0.30 & 0.36 & 0.60 & 0.24 & 0.48 \\
\hline
Time / run (min) & 0.67 & 2.32 & 2.95 & 3.51 & 4.08 & 4.63 & 5.15 & 5.61 & 6.41 & 6.83 \\
\hline
\end{tabular}
}
\end{center}
\vspace{-5mm}
\caption{Comparison of the $\zeta_0$ for the Merton problem and the values 
found using the Monte Carlo method for different values of the dimension 
parameter $K$. The number of Monte Carlo paths each time was equal to
 $\boldsymbol{M = 1000}$. For each set of simulated Monte Carlo paths,
  we find the optimal implied value of $\zeta_0$. We then take the average 
  as the estimate, and calculate its standard deviation.
Here we take $r = 0.05$, $\rho = 0.03$, $R = 3$, $w_0 = 1$, $a = 1$, 
$b = 1$, $N = 100$, $dt = 0.05$. The parameters $\mu$ and $\sigma$ were generated randomly: $\mu$ had a $U[10\%, 50\%$ distribution, once the entries of $\sigma$ were drawn from $U[-1, 1]$ until the resulting matrix was positive definite.}
\label{table3}
\end{table}


\begin{table}
\begin{center}
{\footnotesize
\begin{tabular}{ | c | c | c | c | c | c | c | c | c | c | c | }
\hline
& $K = 1$ & $K =2$ & $K =3 $ & $K =4$ & $K =5$ & $K =6$ & $K =7$ & $K =8$ & $K =9$ & $K =10$ \\
\hline
Merton & 10.10 & 9.67 & 9.34 & 8.63 & 8.35 & 8.13 & 7.33 & 7.10 & 6.83 & 6.48\\
\hline
Average($\zeta_0$) & 10.15 & 9.58 & 9.35 & 8.76 & 8.29 & 8.33 & 7.30 & 7.16 & 6.63 & 6.95 \\
\hline
Stdev($\zeta_0$) & 0.04 & 0.08 & 0.06 & 0.08 & 0.08 & 0.12 & 0.15 & 0.19 & 0.12 & 0.49 \\
\hline
Time / run (min) & 6.87 & 22.90 & 28.65 & 34.42 & 40.06 & 46.25 & 52.08 & 57.27& 61.26 & 68.09 \\
\hline
\end{tabular}
}
\end{center}
\vspace{-5mm}
\caption{Comparison of the $\zeta_0$ for the Merton problem and the values
 found using the Monte Carlo method for different values of the dimension 
 parameter $K$. The number of Monte Carlo paths each time was equal to
  $\boldsymbol{M = 10000}$. For each set of simulated Monte Carlo paths, 
  we find the optimal implied value of $\zeta_0$. We then take the 
  average as the estimate, and calculate its standard deviation. 
  Here we take $r = 0.05$, $\rho = 0.03$, $R = 3$, $w_0 = 1$, $a = 1$, 
  $b = 1$, $N = 100$, $dt = 0.05$. The parameters $\mu$ and $\sigma$ were generated randomly: $\mu$ had a $U[10\%, 50\%$ distribution, once the entries of $\sigma$ were drawn from $U[-1, 1]$ until the resulting matrix was positive definite.}
\label{table4}
\end{table}

\subsection{Non-constant relative risk aversion}
The example of the Merton problem has shown us that the Monte Carlo method can handle 
situations where we deal with a multi-dimensional Brownian motion. However, the 
multiplicative scaling property of the CRRA utility function $u$ means that we are 
unable to assess the accuracy in predicting $\theta$. The remarkable accuracy in 
prediction in Figure \ref{fig:Example1} is caused by the fact that $g(t, \zeta, X) = 
\zeta^{1 - 1/R} \tilde{g}(t,X)$, for some function $\tilde{g}$, and the fact 
that the optimal $\theta$ satisfies \eqref{theta1}.

It will therefore be informative the consider an example where the proportion 
of money invested in the risky assets varies with wealth. This can be done, 
although the price to pay is dimensionality. In this Section, we assume that 
the financial market has constant coefficients and that there is only one asset in the market. 

For $R_1 > 1 > R_2 > 0$, we define the agent's marginal utility as
\begin{align}\label{ICRRA}
I(t, y) & = a_1^{1 / R_1} e^{-\rho t / R_1} y^{- 1 / R_1} + a_2^{1 / R_2}
   e^{-\rho t / R_2} y^{- 1 / R_2}, \\
I_\varphi(y) & = b_1^{1 / R_1} y^{-1 / R_1} + b_2^{1 / R_2} y^{-1/R_2}.
\end{align}

What this means is that, for small values of wealth $w$, the agent's relative risk 
aversion is close to $R_1$ and the agent behaves similarly to the Merton investor 
from Section \eqref{Merton} with $R = R_1$, $a = a_1$ and $b = b_1$, and value 
function $V_1(t, w)$. Conversely, the investor for large values of $w$ is less 
risk averse, with risk aversion $R_2$. He behaves like a Merton investor from 
Section \eqref{Merton} with $R = R_2$, $a = a_2$, and $b = b_2$, and value function $V_2(t, w)$.

In dimension one, there are two very effective methods for solving this problem: 
policy improvement and quantisation\footnote{Both of which are difficult to 
generalise to dimensions more than one, though.}. We proceed by briefly describing 
each one of them, and then by comparing their performance with the Monte Carlo 
scheme we proposed earlier. 

\vspace{3mm}

\noindent \textbf{Policy improvement}. We follow the approach described in 
Section $3.4$ of \cite{Rogers:2013}. The HJB equation for our problem is 
\begin{align}\label{HJBCRRA}
0 = \sup_{c, \theta} \left[ U(t, c) + V_t(t, w) + (rw + \theta(\mu - r) - c) 
V_w(t, w) + \frac{1}{2} \theta^2 \sigma^2 V_{ww} (t, w)\right],
\end{align}
and we are given the terminal value 
\begin{align}
V(T, w) = \varphi(w).
\end{align}

Given functions \eqref{ICRRA}, functions $U$ and $\varphi$, although not 
available in closed form, can be found efficiently using binary search.

We therefore give ourselves a grid of time points $0 < t_1 < t_2 < \dots < t_N = T$ 
and a grid of space points $w_1 < w_2 < \dots < w_M$ and we wish to find $V$ 
evaluated at their mesh. 

At the boundaries, we know that the solution resembles the Merton solutions:
\begin{align}\label{boundaryMerton}
V(t, w_1) = V_1(t, w_1), \qquad V(t, w_N) = V_2(t, w_N).
\end{align}

Let $\mathcal{L}(c, \theta,w)$ be a functional acting on smooth test functions $\psi(t, w)$ as
\begin{align}\label{Ldef}
\mathcal{L}(c, \theta) \psi(t, w) = (rw + \theta(\mu -r) - c) \varphi'(t, w) + 
\frac{1}{2} \theta^2 \sigma^2 \psi''(t, w).
\end{align}

Noticing that 
\begin{align}\label{discritizations}
\psi'(t, w) & \approx \frac{\psi(t, w_{i+1}) - \psi(t, w_{i-1})}{\Delta_+ + \Delta_-} \\
\psi''(t, w) & \approx \frac{\Delta_- (\psi(t, w_{i+1}) - \psi(t, w_i))
 - \Delta_+(\psi(t, w_i) - \psi(t, w_{i-1}))}{\Delta_+ \Delta_- (\Delta_+ + \Delta_-)},
\end{align}
where $\Delta_+ = w_{i+1} - w_i$ and $\Delta_- = w_i - w_{i-1}$, it is possible 
to approximate $\mathcal{L}$ acting on $\psi(t, \cdot)$ by a sparse triagonal 
matrix $L(c, \theta)$ acting on a column vector $\psi(t, w_i), i = 2, \dots M-1$, 
using approximations \eqref{discritizations} plugged into \eqref{Ldef}\footnote{Where 
we consider $w = (w_1, w_2, \dots, w_M)^T$ as a column vector, with the
 corresponding controls $(c_1, c_2, \dots, c_M)^T$ and $(\theta_1, \theta_2, \dots, \theta_M)^T$.}.

We now discretize the differential operator appearing in the HJB equation
 \eqref{HJBCRRA} on the chosen time and space grid. By letting $V_i^n = V(t_n, w_i)$
  and $V^n = (V_i^n)_{i = 1, 2, \dots, M}$, we obtain:
\begin{align}\label{discreteHJB}
0 & = \sup_{c, \theta} \Big[ \frac{V^{n+1} - V^n}{t_{n+1} - t_n}+ 
\alpha( L(c_n, \theta_n) V^n + U(t_n, \cdot, c_n)) \notag
 \\ 
 & \qquad\qquad + (1-\alpha) (L(c_{n+1}, \theta_{n+1})V^{n+1} 
 + U(t_{n+1}, \cdot, c_{n+1}))\Big].
\end{align}
We took $\alpha = 0.5$, giving  the Crank-Nicholson method.
We define $L$ to act on the boundary points $w_1$ and $w_M$ in such a way 
that \eqref{discreteHJB} yields boundary solutions given by \eqref{boundaryMerton}.

Given $(c, \theta)$, \eqref{discreteHJB} is then a sparse set of linear
 equations which we solve for $V$. We then improve on $(c, \theta)$ by 
 maximisation in \eqref{discreteHJB}, given the found $V$. We iterate 
 the process until convergence.

\begin{figure}
\begin{center}
\includegraphics [angle=90,height= 20cm, width=15cm]{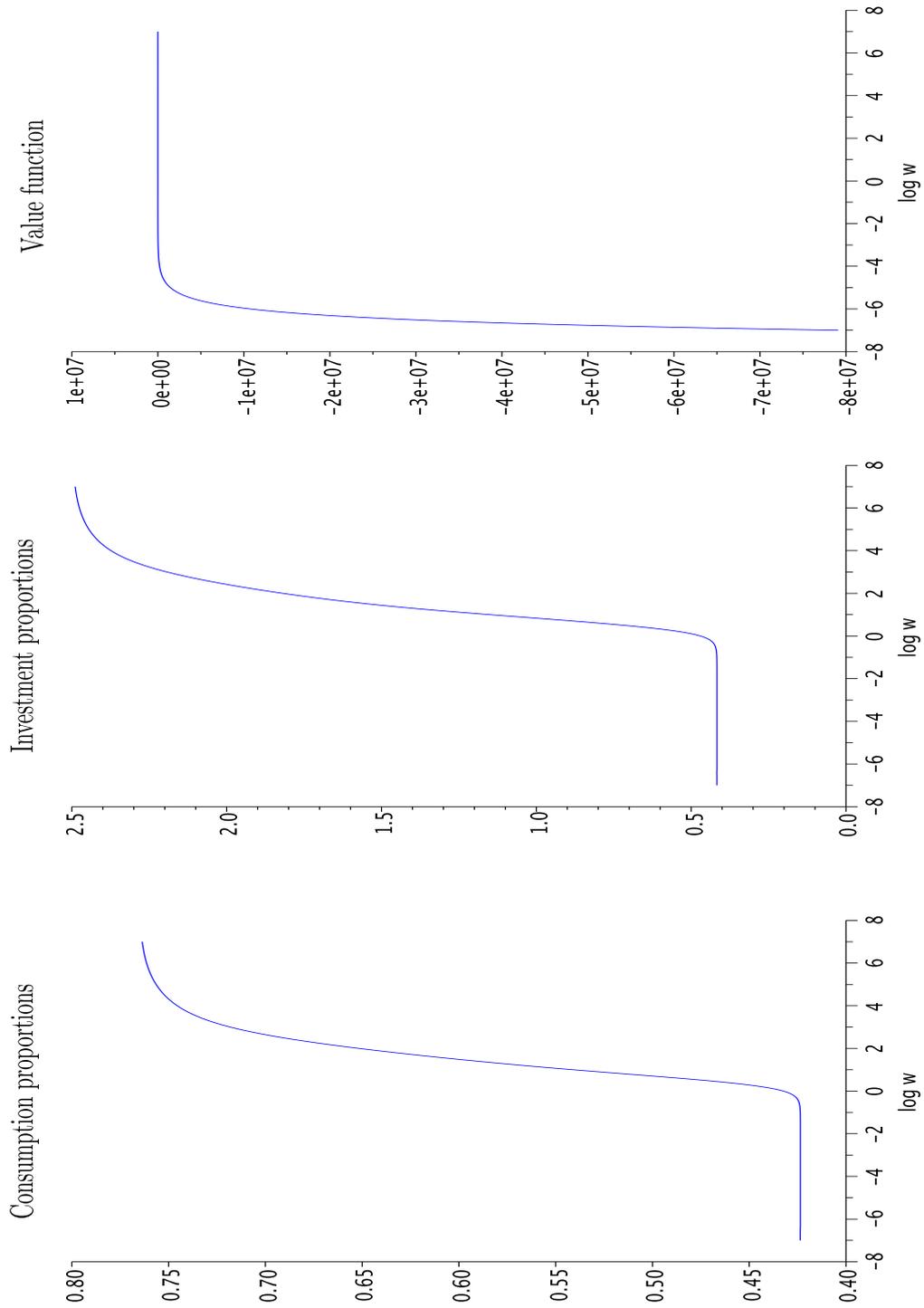} 
\caption{\textbf{Value function found using policy improvement algorithm}. 
Here we take $w_0 = 2$, $\mu = 0.10$, $\sigma = 0.20$, $r = 0.05$,
 $\rho = 0.03$, $a_1 = 10$, $a_2 = 20$, $b_1 = 30$, $b_2 = 10$, $R_1 = 3$,
  $R_2 = 0.5$, $T = 1$, $N = 100$.}
\label{fig:Example2V}
\end{center}
\end{figure}

Figure \ref{fig:Example2V} shows the results of the policy improvement algorithm for $t=0$.
 As we see, we were able to recover the whole value function using the method 
 described above. It is worth pointing out, though, that the method is tricky 
 to implement even in one dimension, and higher dimensions are almost certainly
  out of question. However, once $V$ has been found in one dimension, working 
  out the optimal consumption and investment around a sample path of Brownian 
  motion are immediate. 

\vspace{3mm}


\noindent \textbf{Quantization}. We proceed to a method which builds on 
the observations from Section \ref{sec:problem}, but avoids using the 
Monte Carlo method for approximating the expectation operator in \eqref{gdef}.
 Instead, quantisation proposes approximating the expectation of the 
 Brownian functional by a deterministic sum. Here we follow the details 
 from the website \cite{quantization} and related papers \cite{Pages:2003} and \cite{Corlay:2011}.
The idea is to use the Karhunen-Loeve expansion of Brownian motion $(W_t)_{0 \leq t \leq T}$: 
\begin{align}\label{KLexpansion}
W_t = \sum_{k=1}^\infty \xi_n e_n(t),
\end{align}
where $\left( \xi_n \right)_{n \geq 1} \sim N(0, \lambda_n)$ is a 
sequence of independent normal random variables with variance $\lambda_n$.
 Here the decomposition functions are
\begin{align}
e_n(t) & = \sqrt{\frac{2}{T}} \sin \left( \frac{\pi t}{T} \left(n - \frac{1}{2}\right)\right), \\
\lambda_n & = \left( \frac{T}{\pi(n - \frac{1}{2})}\right)^2.
\end{align}
Brownian motion $W$ is then firstly approximated by choosing the first $d$ 
terms in the sum \eqref{KLexpansion}. We can then think of 
$\xi = (\xi_n)_{n = 1, 2, \dots, d}$ and $e(t) = (e_n(t))_{n = 1, 2, \dots, d}$ 
as $d$-dimensional vectors, and the Brownian motion as being approximated by the dot product
\begin{align}
W_t \approx \xi \cdot e(t)
\end{align}
We then quantise the random $d$-dimensional vector $\xi$ by a random variable 
$X$ taking $n$ distinct values $x_1, x_2, \dots, x_n \in \mathbb{R}^d$ with 
respective probabilities $p_1, p_2, \dots, p_n$, and giving us the final approximation
\begin{align}
W_t \approx X \cdot e(t).
\end{align}
Now, if we need to calculate an expected value of a functional 
\begin{align}\label{quantfunctional}
\mathbb{E}\left[  \int_0^T f(t, W_t) dt + F(W_T) \right],
\end{align}
we can now approximate it by a deterministic sum
\begin{align}
 \sum_{k=1}^n p_k \left[ \int_0^T f(t, x_k \cdot e(t)) dt + F(x_k \cdot e(t)) \right].
\end{align}

The effectiveness of this application depends on the number of terms $n$ taken 
in the expansion \eqref{KLexpansion}, as well as the placing of the points and 
weights $x_i$ and $p_i$. 
Files of the points and weights for many different values of $n$ and for 
dimension up to 10 may be freely downloaded from the website
 \cite{quantization}. These points and weights are optimal quantizations
 of the standard Gaussian distribution, in a sense explained in detail there.
 For a chosen number of $n$, we can therefore load up the 
optimal  $x_i$ and $p_i$ from these files. For our runs, we use $n = 10160$.

The important thing is that, for the  current problem, the calculations we need 
to perform are of the particular form \eqref{quantfunctional}. Indeed, in a 
complete market with one asset, we have 
\begin{align}
g(t, \zeta) = \mathbb{E} \left[ \int_t^T \tilde{U}(t, \zeta_s) ds + 
\tilde{\varphi}(\zeta_T) \Big| \zeta_t = \zeta \right],
\end{align}
compare it with \eqref{gdef}. Here $\zeta$ has a closed-form expression 
\begin{align}
\zeta_s = \zeta_t \exp \left[-\kappa (W_s - W_t) - (r + \frac{1}{2} \kappa^2)(s-t) 
\right] \text{ for } s \geq t,
\end{align}
which is of the required form \eqref{quantfunctional}. 

Having laid out the problem setup and the accurate numerical methods for solving
 the problem, we now show the numerical results of our calculations. 

\vspace{3mm}

\begin{figure}
\begin{center}
\includegraphics [angle=90,height= 20cm, width=15cm]{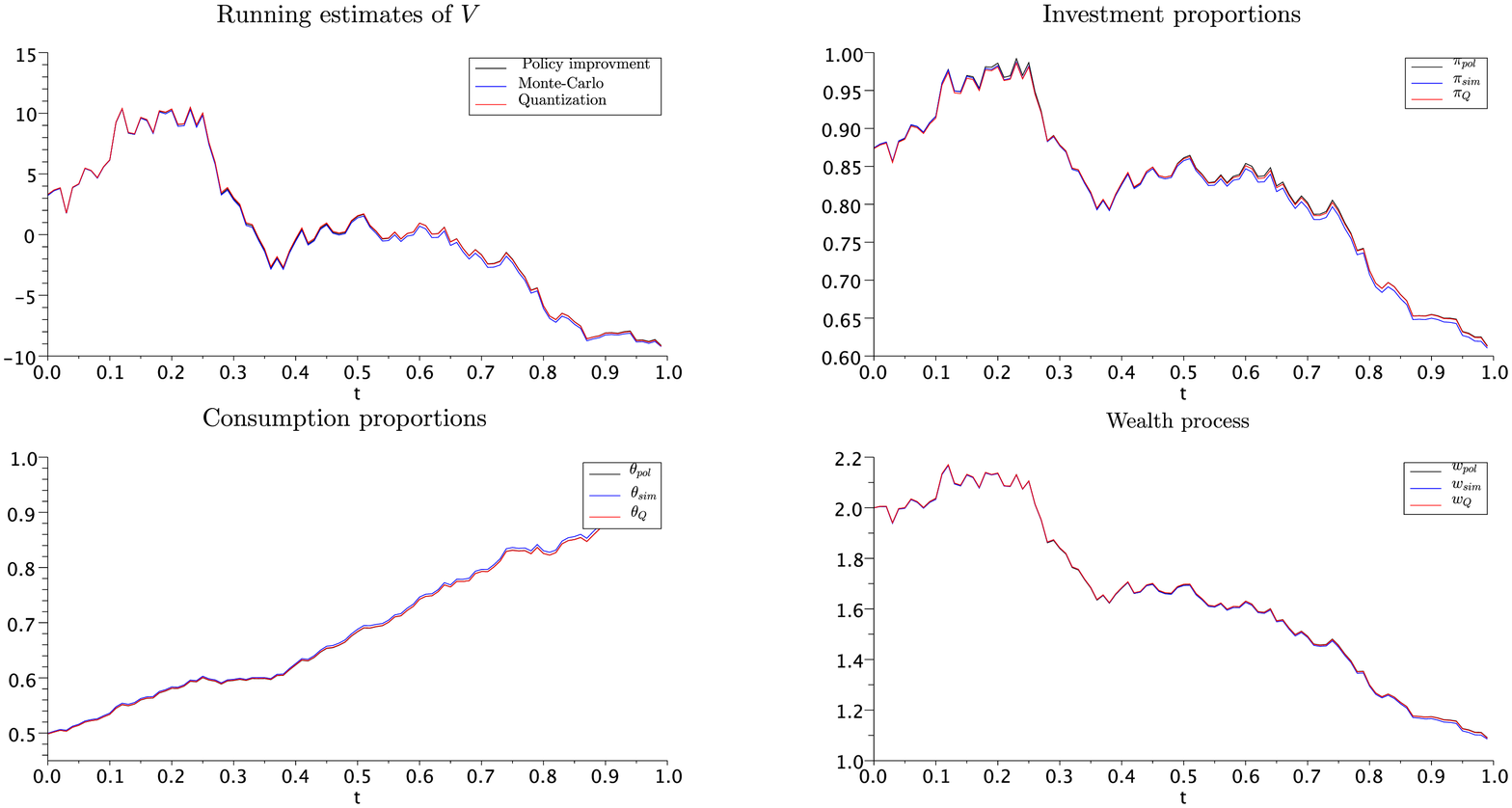} 
\caption{\textbf{Comparison of different methods for the non-constant
relative risk
 aversion example}. Here we take $w_0 = 2$, $\mu = 0.10$, $\sigma = 0.20$, 
 $r = 0.05$, $\rho = 0.03$, $a_1 = 10$, $a_2 = 20$, $b_1 = 30$, $b_2 = 10$,
 $R_1 = 3$, $R_2 = 0.5$, $T = 1$, $N = 100$. The number of Monte Carlo
  paths we took is $M = 10000$.}
\label{fig:Example2}
\end{center}
\end{figure}

\noindent \textbf{Comparison of the methods}. Figure \ref{fig:Example2} shows 
the results of the simulation runs. It is clear that all the methods proposed
 give virtually the same answers; with Monte Carlo being only away from the
  two benchmark methods of policy improvement and quantisation. The most
   reassuring message here is that the Monte Carlo methodology also does
    a very good job on approximating the investment proportions for the 
    problem as in \eqref{theta1} and \eqref{copt}. This is 
    the part the the Merton problem example was unable to reveal
     due to the special structure.

The time taken to get the answer for the policy improvement was approximately
 $10$ minutes, most of which was taken on the calculation of the value 
 function (evaluating the solution along a chosen path is extremely fast). 
 In comparison, quantisation has taken roughly $4$ minutes, and Monte Carlo took $8$ minutes. 

Of course, each of the methods has their costs and benefits. The value function 
takes a time-investment at the start, but is very fast regardless of how many 
sample paths we would like to evaluate. This is not the case for quantisation 
and the Monte Carlo method. Quantisation is the overall speed-winner here, 
however we must remember that this is mainly due to the preloaded files which
 we used to quantise the Brownian motion. 

Overall, we conclude that the Monte Carlo method performs very well on the
 complete market problems, as it should. After all, as mentioned before, the 
 only errors we are incurring are numerical: approximating the derivatives and
  the expectation operator. With sufficient computational power, these should 
  be possible to be made small.

\subsection{Incomplete markets driven by a diffusion}
Finally, we consider an example where no benchmark methods are available, and 
the bounds derived in \eqref{bounds} are the only sensible indicator for how 
well our method is doing. We consider an example that is as challenging as 
possible: an incomplete market driven by a diffusion. 

\begin{figure}
\begin{center}
\includegraphics [angle=90,height= 16cm, width=15cm]{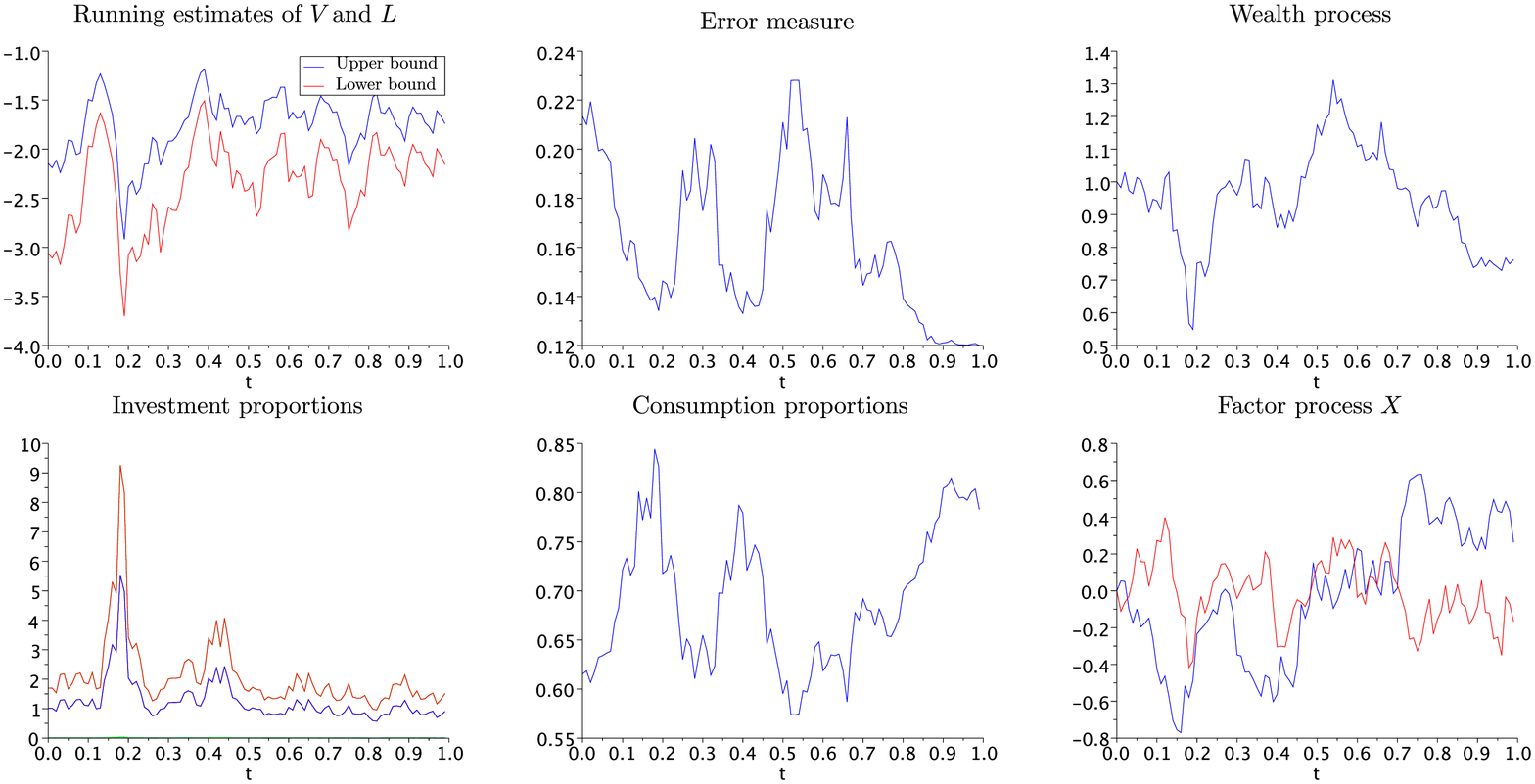} 
\caption{\textbf{An incomplete market driven by a stochastic factor}. 
Here we take an incomplete market $5$-dimensional Brownian motion with $4$
 independent assets and $M=1000$ Monte Carlo paths in the approximation of $g$. 
 We start with $w_0 = 1$,  $\rho = 0.03$, $T = 1$, $N=100$. We take the
  CRRA utility function with parameters $a = 1$, $b =2$, $R=3$. $X$ is 
  taken to be an OU process with randomly-generated volatility matrix 
  and mean-reverting drift. The market interest rate $r$ and $\mu$ are 
  constant and randomly generated. Market volatility $\sigma$ is a 
  random $4 \times 5$ matrix multiplied by a stochastic scaling 
  factor $1 + \exp(- \wun \cdot X_t)$. The randomisation is done by drawing relevant parameters from $U[-1, 1]$ distribution via Gibbs-sampling until the regularity conditions imposed by the paper are met (i.e. $\mu \geq r \geq 0$, $\sigma$ has rank 4 and $\sigma^T \sigma$ is invertible).
  The 
  top panels represent the running upper and lower bounds on the
   objective as defined in \eqref{bounds}, error rate as in \eqref{error}, 
   and the corresponding wealth process from \eqref{dual1}. The bottom 
   panel represents the investment and consumption proportions,
   together with the first two components of the factor process $X$.}
\label{fig:K5M10000}
\end{center}
\end{figure}

As a specific example, we consider a market composed of $4$ stocks driven by a
 $5$-dimensional Brownian motion. The same Brownian motion drives the $5$-dimensional
  factor process $X$, which we assume to a be an OU process with the mean-reversion
   and volatility parameters generated randomly. We take a CRRA utility function,
    with a number of Monte Carlo paths being equal to $M = 1000$. The results
     of the optimisation run are depicted on Figure \ref{fig:K5M10000}. The run 
     time here took $23h$. The details regarding the 
     parameters are displayed below the panel. 

As we can see from the first two panels on the top, the upper and lower bounds 
stay reasonably close during the sample runs, with the error measure defined 
in \eqref{error} between $12\%$ and $22\%$, and generally decreasing as we near the end of investment. 

This is a positive result, especially in light of the dimensionality of the problem.
 Notice that the market is incomplete, and that the value function for this problem
  would need to be $7$-dimensional ($1$ dimension for wealth, $1$ for time, and $5$ 
  for the factor process $X$). Hence, any other method for approaching this problem 
  would really struggle. 

We could of course try to improve on the performance of this algorithm. We
 lose efficiency when we use the approximation of $\kappa$ in
  \eqref{kappa_pi}, and also when we truncate the expression \eqref{theta_star} for
  $\theta$.  
However, our main goal of the paper has already been achieved here: we have illustrated 
how to use our method on a very difficult problem, and derived satisfactory bounds on the efficiency. 

\section{Conclusions}\label{sec:conclusions}
This paper presented an effective methodology for tackling optimal investment problems 
in incomplete markets driven by a Brownian diffusion. We were able to derive a generic
 methodology for numerically tackling these problems by taking some convenient mode
 realisation of the market. Secondly, we settle for suboptimal controls which are 
 close to the optimal control. 

These assumptions are not a weakness of the method; they are rather a necessary cost needed 
to be taken if we want to get concrete investment advise in a general setup. After all, 
they let us derive what we really need in practise: a method for finding a good investment 
strategy when faced by a particular realisation of the market!

We have also illustrated the effectiveness of the method in a variety of contexts.
 For the problems where other reliable numerical techniques are available, we showed
  our method does just as good. For a very complex multi-dimensional problem concluding 
  Section \ref{sec:numerics}, we have showed that the investment errors can be kept
  satisfactory low. No other method was able to provide even estimates of the solutions
   in this context. This proves the effectiveness of the method and shows that it has 
   a potential of giving what we really need: concrete investment prescriptions
    facing a particular market environment.


\pagebreak
  
\bibliography{Duality}

\end{document}